\documentclass[10pt,runningheads]{llncs}

\PassOptionsToPackage{final}{showkeys}

\usepackage{mystyle}
\usepackage{tpic}
\usetikzlibrary{backgrounds,arrows}
\usepackage{pgfplots}
\usepackage{pgfplotstable}
\usepackage{tikz}
\usepackage{booktabs} 

\begin{document}

\title{DeepBundle: Fiber Bundle Parcellation with Graph Convolution Neural Networks}
\titlerunning{DeepBundle}

\authorrunning{F.~Liu, J.~Feng, G.~Chen, Y.~Wu, Y.~Hong, P.-T.~Yap, D.~Shen}   
\author{Feihong Liu\inst{1,2} \and Jun Feng\inst{1,*} \and Geng Chen\inst{2} \and Ye Wu\inst{2} \and Yoonmi Hong\inst{2} \and  Pew-Thian~Yap\inst{2,*} \and Dinggang Shen\inst{2,*}}
\institute{      School of Information Science and Technology, Northwest University, Xi'an, China \and	
                 Department of Radiology and Biomedical Research Imaging Center (BRIC), University of North Carolina at Chapel Hill, Chapel Hill, U.S.A. \\
    	\Letter~\email{fengjun@nwu.edu.cn},
        \Letter~\email{ptyap@med.unc.edu},
        \Letter~\email{dgshen@med.unc.edu}
}

\maketitle

\begin{abstract}

Parcellation of whole-brain tractography streamlines is an important step for tract-based analysis of brain white matter microstructure. Existing fiber parcellation approaches rely on accurate registration between an atlas and the tractograms of an individual, however, due to large individual differences, accurate registration is hard to guarantee in practice. To resolve this issue, we propose a novel deep learning method, called DeepBundle, for registration-free fiber parcellation. Our method utilizes graph convolution neural networks (GCNNs) to predict the parcellation label of each fiber tract. GCNNs are capable of extracting the geometric features of each fiber tract and harnessing the resulting features for accurate fiber parcellation and ultimately avoiding the use of atlases and any registration method. We evaluate DeepBundle using data from the Human Connectome Project. Experimental results demonstrate the advantages of DeepBundle and suggest that the geometric features extracted from each fiber tract can be used to effectively parcellate the fiber tracts. 

\end{abstract}

\section{Introduction}

Diffusion MRI provides valuable insights into the $3$D geometric structure of brain neural fiber tracts, allowing tract-based analysis (TBA) of brain white matter microstructure in vivo \cite{ciccarelli2008diffusion}. TBA usually focuses on specific fiber bundles that are extracted from the whole-brain tractograms. Therefore, effective methods for segmenting whole-brain tractograms into fiber bundles of interest are desirable.

Existing fiber parcellation approaches can be divided into two categories: ROI-based and streamline-based. 
ROI-based approaches~\cite{yendiki2011automated, wassermann2016white} first parcellate the brain surface based on an atlas and then use the parcellation ROIs to extract different fiber bundles. Streamline-based approaches~\cite{garyfallidis2015robust,zhang2018anatomically} employ streamline registration methods to directly transfer the fiber parcellation information from an atlas to tractograms of an individual. These methods are directly affected by registration accuracy, which is, however, hard to guarantee in practice due to factors such as inter-subject variability, noise, and artifacts. Moreover, the registration procedure is usually time-consuming, making it unsuitable for large-scale studies and real-time applications. Recently, Wasserthal et al. \cite{wasserthal2018tractseg} proposed to predict the binary mask of a fiber bundle from the whole brain fiber peaks using convolutional neural networks. This method avoids the registration procedure but is limited to generating a binary tract segmentation mask rather than the parcellation labels of the fiber tracts. In addition, this method utilizes the fiber orientation information at each voxel rather than the actual geometries of the fiber tracts.

In this work, we propose a deep-learning approach, called DeepBundle, to parcellate whole-brain tractograms without the time-consuming registration. Specifically, we view the coordinates of the points on each fiber tract as functions defined on a graph. The point coordinates are then fed to a graph convolutional neural networks (GCNN) to extract the latent geometric features of fiber tracts for bundle recognition. Our network is trained end-to-end with the point coordinates as inputs and parcellation labels as outputs, thus avoiding having to manually craft features for the purpose of bundle parcellation. During the testing stage, the point coordinates extracted from fiber tracts of the whole brain are directly fed to the trained networks, thus avoiding the use of atlases and any registration method. Extensive experiments performed using data from the Human Connectome Project (HCP) indicate that DeepBundle yields fiber parcellation labels with remarkably improved accuracy, confirming that the geometric features extracted by GCNNs are effective for fiber bundle parcellation.

\section{Methods}

In this section, we will first show how fiber streamlines can be represented using graphs, and then, introduce the theory of spectral graph convolution and graph pooling. Finally, we will describe our network in detail.

\subsection{Graph Representation of Fiber Tracts}

Considering an undirected graph as $\mathcal{G}=(\mathcal{V},\mathcal{E},\mathcal{W})$, $\mathcal{V}=\left\{v_1, \cdots, v_n \right\}$ is a set of $n$ vertices, $\mathcal{E}\subseteq \mathcal{V} \times \mathcal{V} $ is the edge set, and $\mathcal{W}=\left(w_{\left(i,j\right)}\right)$ is the $n\times n$ adjacency matrix, which is symmetric, i.e., $w_{\left(i,j\right)}=w_{\left(j,i\right)}$.
We uniformly sample $n$ points on each fiber tract, and thus this discrete fiber tract can be represented by a line-type graph, where the edge weights are given by  
\begin{equation}
w_{i,j}=\begin{cases}
1, & \text{if}~i,j~\text{are connected},\\
0, & \text{otherwise}.
\end{cases}
\end{equation}
The graph $\mathcal{G}$ now encodes the geometric relationships between sampling points on the fiber tract~\cite{bronstein2017geometric}. We then view the point coordinates extracted from a fiber tract as graph-structured data.

\subsection{Spectral Graph Convolution}
To extract the underlying geometry-invariance features of each fiber tract, we employ spectral graph convolution~\cite{bruna2014spectral}, which generalizes conventional convolution in Euclidean space using graph Fourier transformation. The transformation relies on the eigendecomposition of the graph Laplacian $\Delta$, which is defined as  
\begin{equation}
    \Delta = \varPhi \Lambda \varPhi^{\text{T}} = I_n - \mathcal{D}^{-1/2}\mathcal{W}\mathcal{D}^{-1/2},
\label{e1}
\end{equation}
where $\varPhi = \left( \varphi_1, \cdots, \varphi_n \right) $ is a matrix of orthonormal eigenvectors ($\varPhi^{\text{T}}\varPhi=I_n$),  $\varLambda=\text{diag}(\lambda_1,\cdots,\lambda_n)$ is a diagonal matrix of corresponding eigenvalues, $I_n$ is an identity matrix, and $\mathcal{D}=\sum_{j\ne i} w_{(i,j)}$ is the degree matrix. Such eigenvectors $(\varphi_1, \cdots, \varphi_n )$ can be interpreted as the Fourier bases, and $\varPhi^T$ is used to transform the features from the spatial domain to spectral domain. The graph Fourier transformation is formulated by
\begin{equation}
\text{f}*\text{g} = \varPhi \left(\varPhi^{\text{T}}\text{g}\,\right)\odot\left(\varPhi^{\text{T}}\text{f}\,\right)=\varPhi \text{diag}\left(\hat{g}_1,\cdots, \hat{g}_n\right)\hat{\text{f}},
\label{e2}
\end{equation}
where $\text{f}=(f_1, \cdots, f_n)^{\text{T}}$ is the input signal, which denotes the geometric features of one streamline on the vertices of graph $\mathcal{G}$. Its Fourier transformation is given by $\hat{\text{f}}=\left(\varPhi^{\text{T}}\text{f}\,\right)$;
$\text{g}$ is a convolutional filter in the spatial domain, and the spectral convolution can be defined as element-wise product $\left(\varPhi^{\text{T}}\text{g}\right)\odot\left(\varPhi^{\text{T}}\text{f}\,\right)$, which can also be written by $\text{diag}\left(\hat{g}_1, \cdots, \hat{g}_n\right)\hat{\text{f}}$. Hereby, $\text{diag}\left(\hat{g}_1, \cdots, \hat{g}_n\right)$ is the corresponding convolutional filter in the spectral domain.

Utilizing the spectral definition of the convolution, GCNNs generalize CNNs to graphs. The $\ell$-th spectral convolution layer can be written by
\begin{equation}
\text{f}_{k}^{(\ell+1)} = \xi\left(\varPhi\sum_{k'=1}^{p}\text{diag}\left(\hat{g}_{(k,k',1)}^{(\ell)}, \cdots, \hat{g}_{(k,k',n)}^{(\ell)}\right)\varPhi^{T}\text{f}_{k'}^{(\ell)}\right), 
\label{e3}
\end{equation}
where $\text{F}^{(\ell)} =\left(\text{f}_{1}^{(\ell)},\cdots,\text{f}_{p}^{(\ell)} \right)$ and $\text{F}^{(\ell+1)} =\left(\text{f}_{1}^{(\ell+1)},\cdots,\text{f}_{q}^{(\ell+1)} \right)$ are the $n \times p$ and $n \times q$ matrices, representing the input and output features of $\ell$-th spectral convolution layer; $\text{f}_{k'}^{(\ell)}$ and $\text{f}_{k}^{(\ell+1)}$ denote the $k'$-th column of the input matrix and the $k$-th column of the output matrix. $\varPhi=\left(\varphi_{1},\cdots, \varphi_{n}\right)$ is an $n\times n$ matrix, and $\text{diag}\left(\hat{g}_{(k,k',1)}^{(\ell)},\cdots,\hat{g}_{(k,k',n)}^{(\ell)}\right)$ is a $n\times n$ diagonal matrix which denotes learnable filters in the spectral domain; and $\xi$ denotes the nonlinearity, e.g.,\,ReLU. 


\subsection{Fast Graph Pooling}
\begin{figure}[!t]
	\centering
	\includegraphics[width=0.98\textwidth]{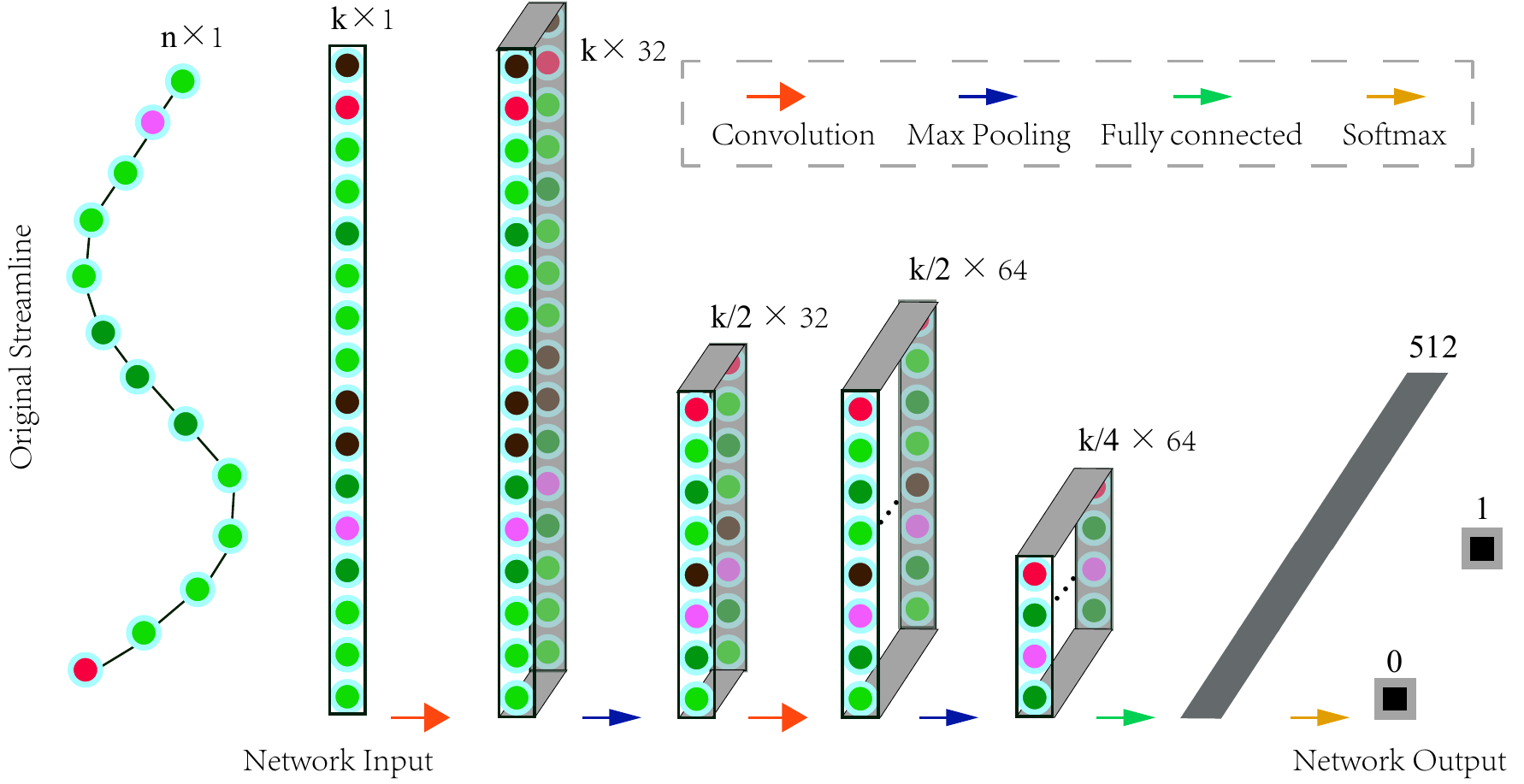}
	\caption{The architecture of the employed spectral GCNNs. Red arrows denote the spectral convolution layers, blue arrows denote the fast max-pooling layers, a green arrow denotes one fully connected layer, and the yellow arrow denotes the softmax layer.}
	\label{fig:net}
\end{figure}

We utilize the Graclus~\cite{dhillon2007weighted,defferrard2016convolutional} to coarsen the input graph into multi-scales, which is similar to the pooling operation in conventional CNNs. In practice, the multi-scale graph is organized as a binary tree, which becomes coarse from the leaf layer to the root layer. When constructing this tree, Graclus introduces fake nodes, rearranges the nodes of the streamline, and thus, we can perform graph pooling using fast $1$D spatial pooling. Cousin nodes in one layer are aggregated into one parent node in the upper layer of the binary tree. 
The pooling process is illustrated in Fig.~\ref{fig:net}.

\subsection{Network Architecture of DeepBundle}
Fig.~\ref{fig:net} illustrates the network architecture of DeepBundle. We employ a spectral GCNNs with the architecture of GC$32$-P$2$-GC$64$-P$2$-FC$512$, where GC$c$ denotes a graph convolution layer which has $c$ filters, P$2$ denotes a graph pooling layer with a factor $2$, and FC$512$ denotes a fully connected layer with $512$ hidden nodes. Each graph convolutional layer is activated by the ReLU function. The last layer is the softmax regression layer, and we employ the cross-entropy loss with an $l_2$ regularization term. We separately trained networks for segmenting different fiber bundles of interest from the whole-brain tracts. 

In Fig.~\ref{fig:net}, the length of the input is increased from $n$ to $k$ due to the introduction of fake nodes (black dots) by Graclus. We employed $32$ and $64$ filters in the first convolution and the second convolution layer respectively. In the pooling layer, two neighbor nodes are aggregated into one.
The input with a length $k$ is reduced to $k/4$ after going through two max-pooling layers. 

\section{Results}
\subsection{Implementation Details}
For evaluation, we used the publicly available HCP fiber tract dataset~\cite{wasserthal2018tract}. 
This dataset contains $105$ subjects. Each subject has $72$ fiber bundles, containing streamlines of  different lengths. We uniformly resampled them to have the same number of points, i.e.,\,$100$ points. We then represented all streamlines using a common graph. In our experiments, we randomly selected $25$ subjects from the database for training, $2$ for validation, and $11$ for testing. In preparing the training data, a positive label is assigned to all streamlines from a bundle of interest. The negative samples are collected by 1) selecting streamlines from spatially neighboring bundles, and 2) randomly selecting an equal number of streamlines from all other bundles. For testing, all streamlines of each testing subject were fed separately into the trained network.

\subsection{Experimental Results}

\begin{figure}[!tb]
	\begin{tabular}{m{0.02\textwidth}m{0.31\textwidth}m{0.01\textwidth}m{0.02\textwidth}m{0.31\textwidth}m{0.01\textwidth}m{0.02\textwidth}m{0.31\textwidth}}
		
		\rotatebox{90}{\centerline{\textbf{CST left}}} &
		\includegraphics[height=0.24\textwidth]{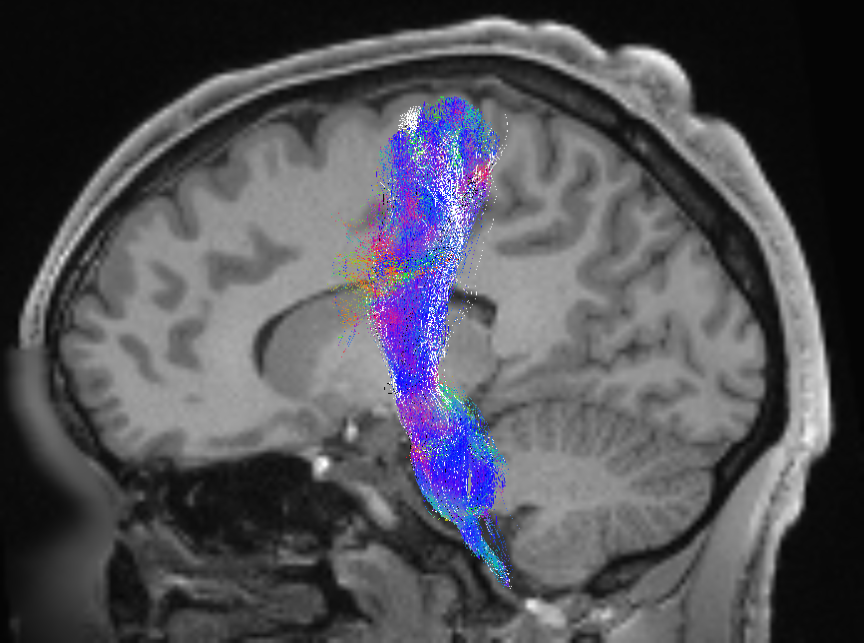}&&
		\rotatebox{90}{\centerline{\textbf{CST right}}} &
		\scalebox{-1}[1]{\includegraphics[height=0.24\textwidth]{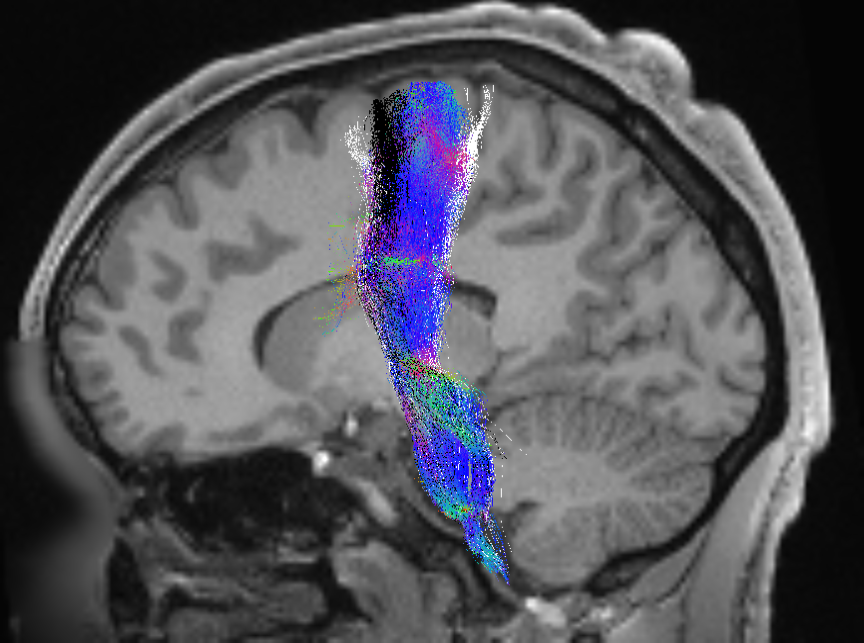}}&&
		\rotatebox{90}{\centerline{\textbf{CA}}} &
		\includegraphics[height=0.24\textwidth]{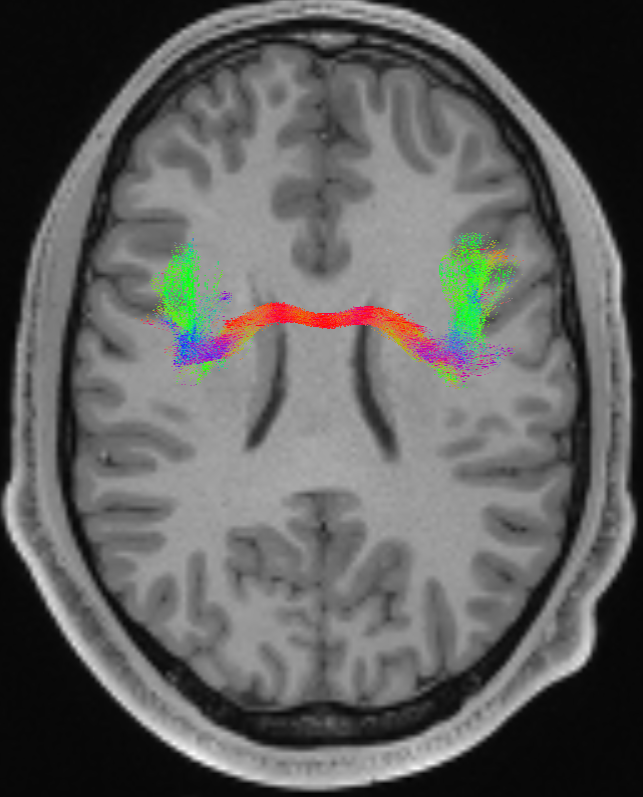}\\
		
		\rotatebox{90}{\centerline{\textbf{UF left}}} &
		\includegraphics[height=0.24\textwidth]{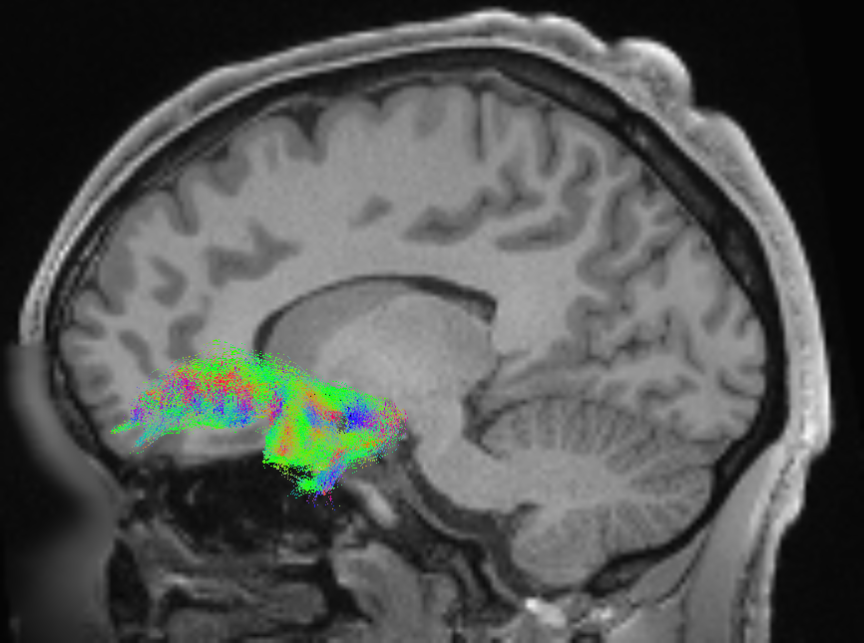}&&
		\rotatebox{90}{\centerline{\textbf{UF right}}} &
		\scalebox{-1}[1]{\includegraphics[height=0.24\textwidth]{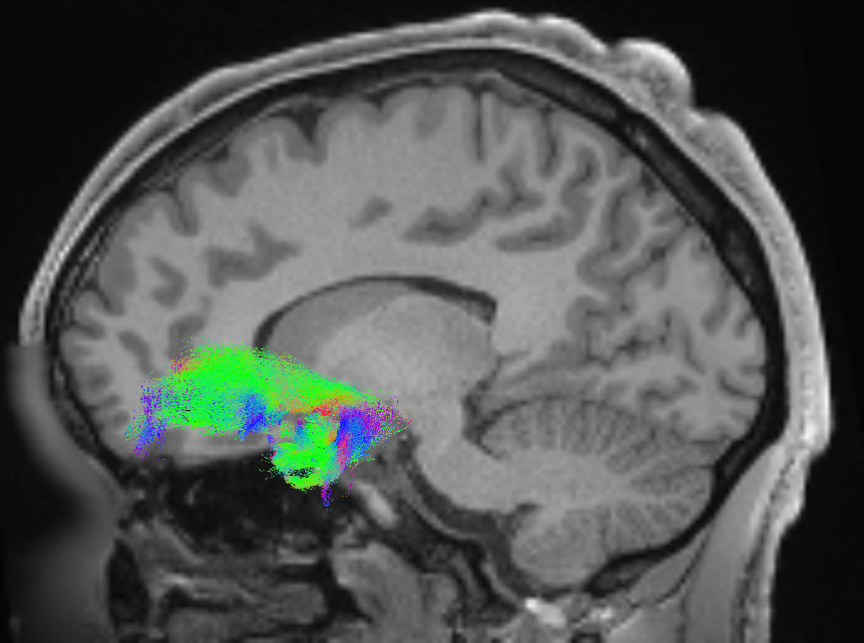}}&&
		\rotatebox{90}{\centerline{\textbf{CC 1}}} &
		\includegraphics[height=0.24\textwidth]{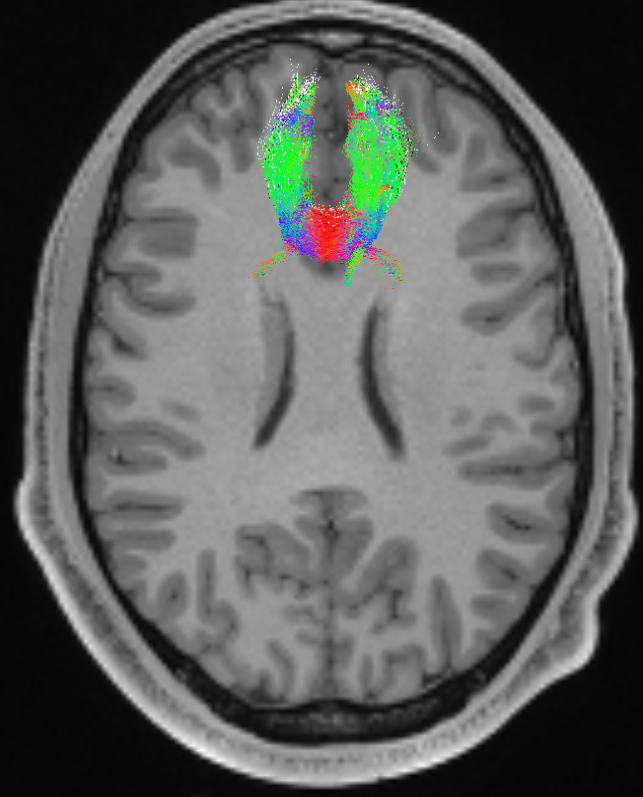}\\
		
		\rotatebox{90}{\centerline{\textbf{FX left}}} &
		\includegraphics[height=0.24\textwidth]{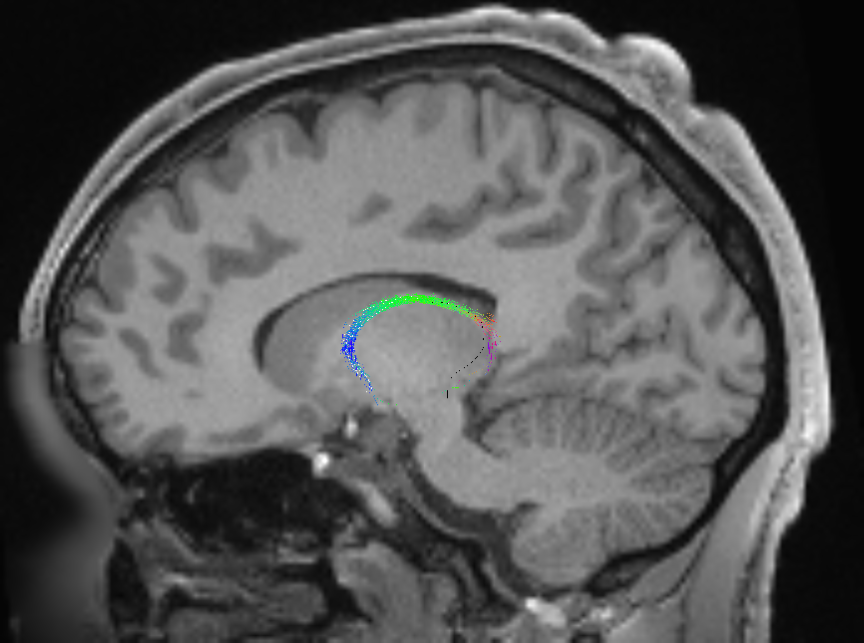}&&
		\rotatebox{90}{\centerline{\textbf{FX right}}} &
		\scalebox{-1}[1]{\includegraphics[height=0.24\textwidth]{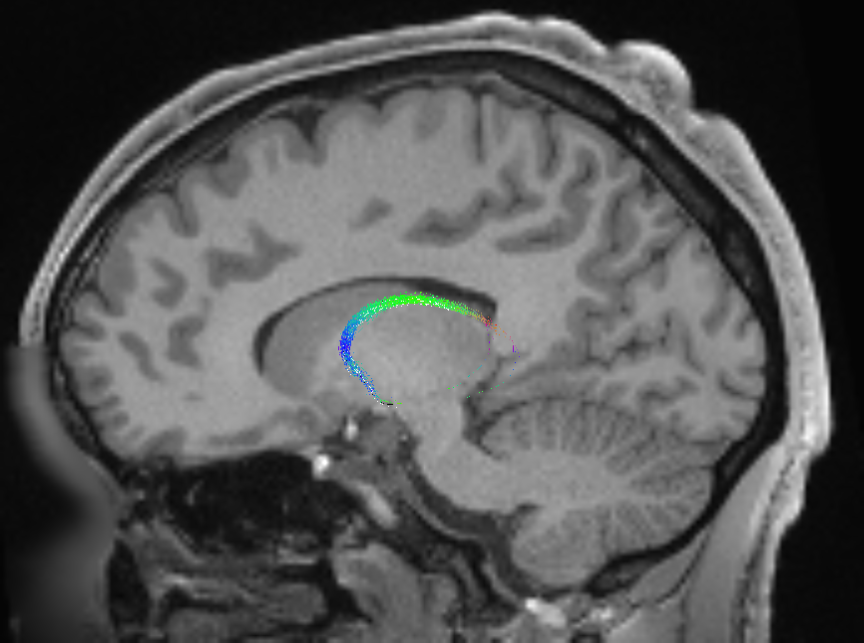}}&&
		\rotatebox{90}{\centerline{\textbf{CC 2}}} &
		\includegraphics[height=0.24\textwidth]{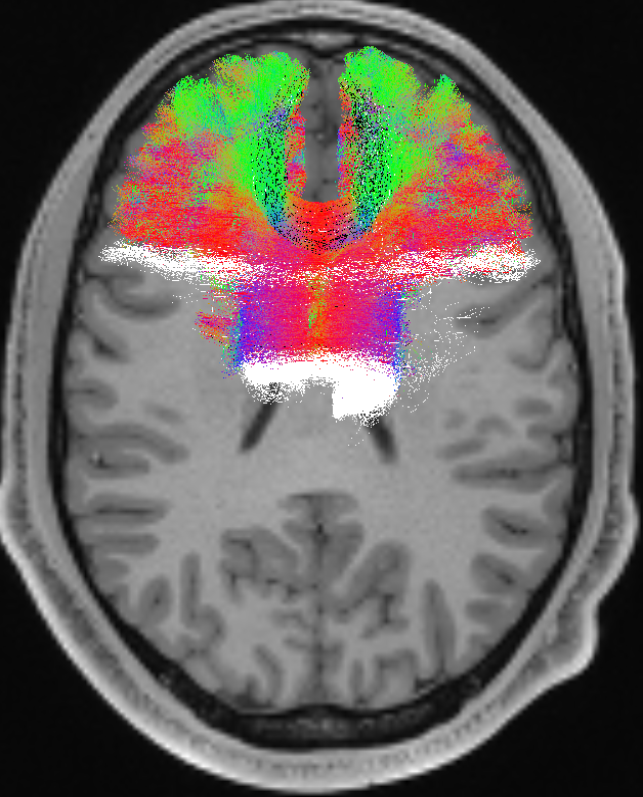}\\
		
		\rotatebox{90}{\centerline{\textbf{AF left}}} &
		\includegraphics[height=0.24\textwidth]{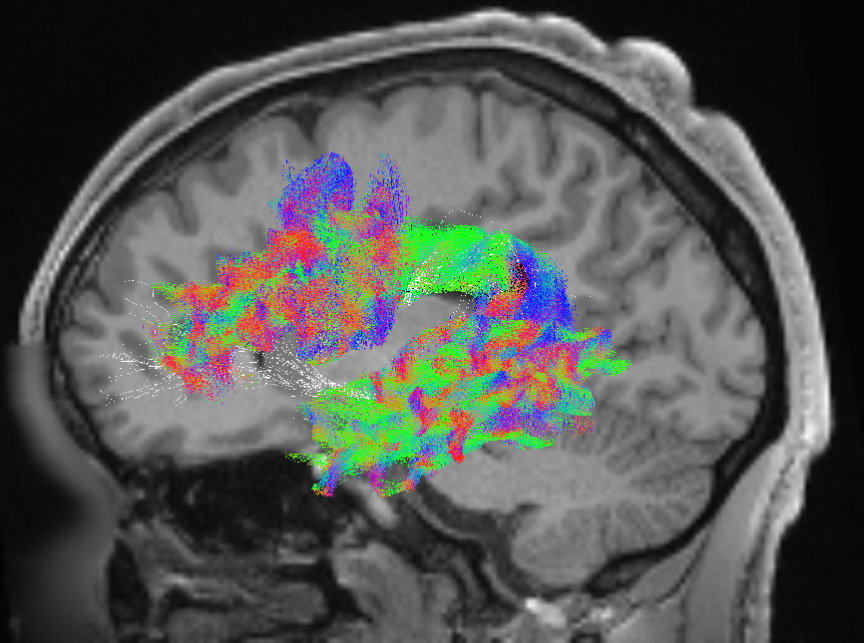}&&
		\rotatebox{90}{\centerline{\textbf{AF right}}} &
		\scalebox{-1}[1]{\includegraphics[height=0.24\textwidth]{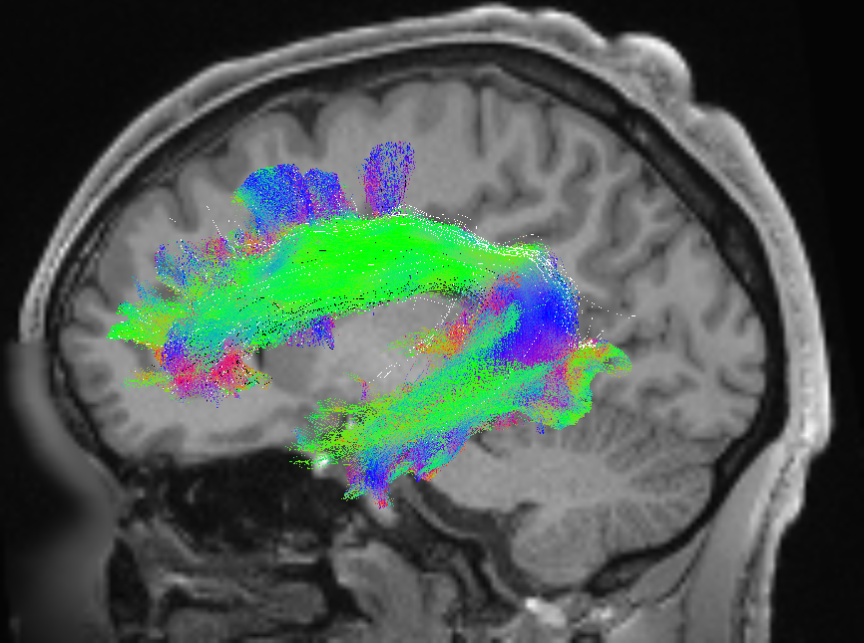}}&&
		\rotatebox{90}{\centerline{\textbf{CC 6}}} &
		\includegraphics[height=0.24\textwidth]{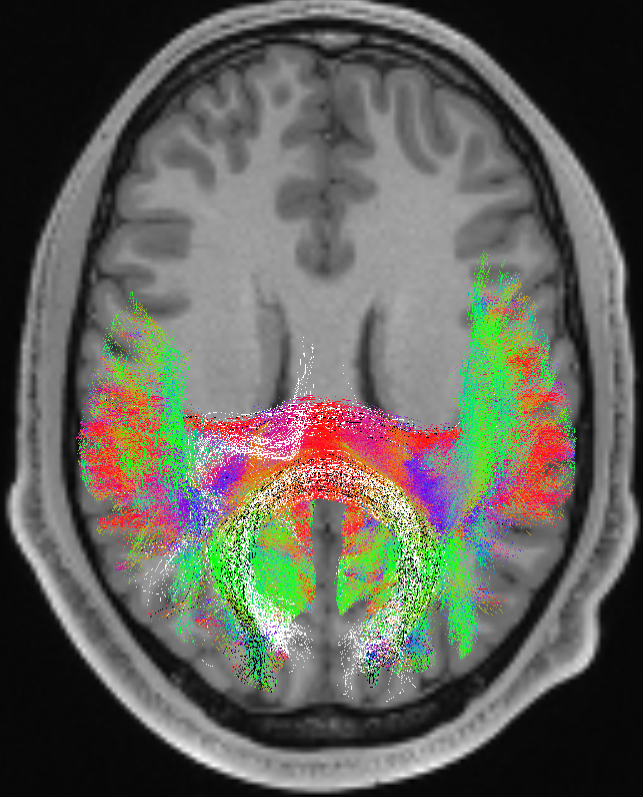}\\
	\end{tabular}
	
	\caption{Qualitative results of $12$ fiber bundles: Corticospinal tract (CST),  Commissure anterior (CA), Corpus callosum (Rostrum (CC $1$), Genu (CC $2$), Isthmus (CC $6$)), Uncinate fascicle (UF), Arcuate fascicle (AF), Fornix (FX). Colored streamlines indicate the true positive (TP) streamlines, while black and white denote the false positive (FP) and false negative (FN) streamlines, respectively. 
  }
	\label{fig:orig}
\end{figure}

Fig.~\ref{fig:orig} shows the fiber parcellation results of one randomly selected subject ($\#623844$). 
$12$ bundles of interest are parcellated from more than one million whole-brain fiber tracts. Although fiber bundles in the left and right hemispheres may share similar shapes, DeepBundle is able to distinguish between them. 
For some small-size bundles, e.g.,\,FX, which have only hundreds of streamlines, DeepBundle consistently gives promising parcellation results. In addition, we show the quantitative parcellation results of the $12$ fiber bundles in Table~\ref{t1}, further confirming the conclusions that can be drawn from Fig.~\ref{fig:orig}.

\begin{table}[bt]
	\centering
	\caption{\label{t1} Classification results for $12$ bundles of subject $\#623844$.}
	\begin{tabular}{m{1.3cm}m{1.55cm}m{1.55cm}m{1.55cm}m{1.55cm}m{1.55cm}m{1.55cm}@{}ccccccc@{}}
		\toprule
		\small{Counts}               & CST left   & CST right  & CC 1       & CC 2       & CC 6       & CA           \\
		\cmidrule(r){2-7}
		TP                & $4755$     & $7726$     & $4419$     & $51967$    & $30082$    & $1330$       \\
		FP                & $244$      & $395$      & $143$      & $1100$     & $602$      & $0$          \\
		FN                & $54$       & $382$      & $29$       & $153$      & $320$      & $0$          \\
		\midrule
		\small{Counts}                  & UF left    & UF right    & AF left    & AF right   & FX left    & FX right   \\
		\cmidrule(r){2-7}
		TP                & $3535$     & $4916$      & $54448$    & $42635$    & $94$       & $123$      \\
		FP                & $0$        & $0$         & $199$      & $495$      & $6$        & $7$        \\
		FN                & $1$        & $3$         & $190$      & $220$      & $1$        & $1$        \\
		\bottomrule
	\end{tabular}
\end{table}

\begin{table}[bt]
	\centering
	\caption{\label{t2} Classification performance (mean $\pm$ standard deviation in $\%$) on $12$ fiber bundles from $11$ subjects.}
	\begin{tabular}{m{1.6cm}m{1.6cm}m{1.6cm}m{1.6cm}m{1.6cm}m{1.6cm}m{1.6cm}@{}ccccccc@{}}
		\toprule
		\            & CST left      & CST right     & CC 1          & CC 2          & CC 6          & CA             \\
		\cmidrule(r){2-7}
		Precision    & $90.5\pm 7.1$ & $91.1\pm 7.9$ & $96.5\pm 2.1$ & $97.6\pm 2.2$ & $95.6\pm 2.3$ & $99.9\pm 0.1$  \\
		Recall       & $88.4\pm 11.6$& $86.6\pm 11.6$& $95.5\pm 4.3$ & $98.4\pm 1.3$ & $98.1\pm 1.8$ & $100 \pm 0.0$  \\	
		\midrule
		\            & UF left      &   UF right     & AF left       & AF right      & FX left       & FX right \\
		\cmidrule(r){2-7}
		Precision    & $99.7\pm 0.6$ & $99.8\pm 0.2$ & $99.2\pm 1.0$ & $99.4\pm 0.6$ & $87.3\pm 12.7$& $90.7\pm 14.1$  \\
		Recall       & $99.2\pm 2.1$ & $99.6\pm 0.4$ & $99.4\pm 0.8$ & $96.5\pm 2.8$ & $97.5\pm 3.9$ & $96.6\pm 3.8$  \\	
		
		\bottomrule
	\end{tabular}
\end{table}

Table~\ref{t2} shows the quantitative results computed across $10$ testing subjects. A large \textit{precision}/\textit{recall} value indicates more accurate fiber parcellation. It can be observed that DeepBundle achieves high \textit{precision} and \textit{recall} values for all bundles of interest, indicating promising fiber parcellation accuracy. It is interesting to note that the recall numbers of left CST and right CST are smaller than the other fiber bundles. This is due to the fact their neighboring bundles, such as POPT and FPT, are very similar to CST in shape. 
Such streamlines induce ambiguities in fiber parcellation, even when performed manually. 
 
We also compared DeepBundle with a popular method called RecoBundles~\cite{garyfallidis2018recognition}. We mapped all parcellated fiber streamlines to created volumetric visitation maps and computed their Dice scores with respect to the gold standard. The results, shown in Table~\ref{t3}, indicate that DeepBundle significantly improves the Dice score and outperforms RecoBundles for all bundles of interest. RecoBundles gives significantly lower Dice scores CA, left FX, and right FX. DeepBundle yields relatively smaller Dice scores for left CST and right CST, but still significantly outperforms RecoBundles.

\begin{table}[bt]
	\centering
	\caption{\label{t3} Dice scores (mean $\pm$ standard deviation in $\%$) of $12$ fiber bundles from $11$ testing subjects.}
	\begin{tabular}{m{2.1cm}m{1.55cm}m{1.55cm}m{1.55cm}m{1.55cm}m{1.55cm}m{1.55cm}@{}ccccccc@{}}
		\toprule
		\                 & CST left         & CST right        & CC 1             & CC 2            & CC 6             & CA           \\
		\cmidrule(r){2-7}
		RecoBundles       & $45.3\pm9.7$     & $40.6\pm6.1$     & $51.3\pm13.2$    & $63.6\pm8.6$   & $60.7\pm6.31$     & $25.2\pm16.1$     \\
		DeepBundle        & $80.7\pm10.1$    & $86.9\pm5.1$     & $93.0\pm3.3$     & $96.8\pm1.2$    & $97.1\pm1.4$     & $99.1\pm1.6$       \\
		\midrule
		\                 & UF left          & UF right         & AF left          & AF right        & FX left          & FX right    \\
		\cmidrule(r){2-7}		
		RecoBundles       & $46.9\pm6.3$     & $52.1\pm7.7$     & $61.3\pm9.1$     & $63.1\pm12.7$   & $8.1\pm2.8$      & $9.5\pm3.6$      \\
		DeepBundle        & $96.4\pm5.1$     & $98.1\pm1.3$     & $95.8\pm3.8$     & $96.2\pm4.0$    & $82.4\pm9.3$     & $86.5\pm8.8$      \\
		\bottomrule
	\end{tabular}
\end{table}

\section{Conclusion}
In this paper, we have proposed a framework for fiber bundle parcellation using a GCNN. Our method directly predicts a tract parcellation label from the point coordinates extracted from a fiber tract. GCNNs are capable of extracting robust geometric features for tract parcellation in an end-to-end manner. Experiments on HCP data demonstrate that our method, DeepBundle, yields promising tract bundle parcellation results with high precision and recall rates. DeepBundle also achieves much higher Dice scores compared with RecoBundles. Our results also suggest that DeepBundle is even capable of effectively parcellating small tract bundles.

\section*{Acknowledgment}
This work was supported in part by NIH grants NS093842 and the Xi'an Science and Technology Project funded by the Xi'an Science and Technology Bureau under Grant 201805060ZD11CG44. 

\bibliographystyle{llncs_splncs}
\bibliography{references}

\begin{thebibliography}{10}

\bibitem{ciccarelli2008diffusion}
Ciccarelli, O., Catani, M., Johansen-Berg, H., Clark, C., Thompson, A.:
\newblock Diffusion-based tractography in neurological disorders: concepts,
  applications, and future developments.
\newblock The Lancet Neurology \textbf{7}(8) (2008)  715--727

\bibitem{yendiki2011automated}
Yendiki, A., Panneck, P., Srinivasan, P., Stevens, A., Z{\"o}llei, L.,
  Augustinack, J., Wang, R., Salat, D., Ehrlich, S., Behrens, T.,  et~al.:
\newblock Automated probabilistic reconstruction of white-matter pathways in
  health and disease using an atlas of the underlying anatomy.
\newblock Frontiers in Neuroinformatics \textbf{5} (2011) ~23

\bibitem{wassermann2016white}
Wassermann, D., Makris, N., Rathi, Y., Shenton, M., Kikinis, R., Kubicki, M.,
  Westin, C.F.:
\newblock The white matter query language: a novel approach for describing
  human white matter anatomy.
\newblock Brain Structure and Function \textbf{221}(9) (2016)  4705--4721

\bibitem{garyfallidis2015robust}
Garyfallidis, E., Ocegueda, O., Wassermann, D., Descoteaux, M.:
\newblock Robust and efficient linear registration of white-matter fascicles in
  the space of streamlines.
\newblock NeuroImage \textbf{117} (2015)  124--140

\bibitem{zhang2018anatomically}
Zhang, F., Wu, Y., Norton, I., Rigolo, L., Rathi, Y., Makris, N., O'Donnell,
  L.J.:
\newblock An anatomically curated fiber clustering white matter atlas for
  consistent white matter tract parcellation across the lifespan.
\newblock Neuroimage \textbf{179} (2018)  429--447

\bibitem{wasserthal2018tractseg}
Wasserthal, J., Neher, P., Maier-Hein, K.H.:
\newblock Tract{S}eg - fast and accurate white matter tract segmentation.
\newblock NeuroImage \textbf{183} (2018)  239--253

\bibitem{bronstein2017geometric}
Bronstein, M.M., Bruna, J., LeCun, Y., Szlam, A., Vandergheynst, P.:
\newblock Geometric deep learning: {G}oing beyond euclidean data.
\newblock IEEE Signal Processing Magazine \textbf{34}(4) (2017)  18--42

\bibitem{bruna2014spectral}
Bruna, J., Zaremba, W., Szlam, A., Lecun, Y.:
\newblock Spectral networks and locally connected networks on graphs.
\newblock In: International Conference on Learning Representations ({ICLR}).
  (2014)

\bibitem{dhillon2007weighted}
Dhillon, I.S., Guan, Y., Kulis, B.:
\newblock Weighted graph cuts without eigenvectors: {A} multilevel approach.
\newblock IEEE Transactions on Pattern Analysis and Machine Intelligence
  {(TPAMI)} \textbf{29}(11) (2007)  1944--1957

\bibitem{defferrard2016convolutional}
Defferrard, M., Bresson, X., Vandergheynst, P.:
\newblock Convolutional neural networks on graphs with fast localized spectral
  filtering.
\newblock In: Advances in Neural Information Processing Systems ({NeurIPS}).
  (2016)  3844--3852

\bibitem{wasserthal2018tract}
Wasserthal, J., Neher, P.F., Maier-Hein, K.H.:
\newblock Tract orientation mapping for bundle-specific tractography.
\newblock In: International Conference on Medical Image Computing and
  Computer-Assisted Intervention ({MICCAI}), Springer (2018)  36--44

\bibitem{garyfallidis2018recognition}
Garyfallidis, E., C{\^o}t{\'e}, M.A., Rheault, F., Sidhu, J., Hau, J., Petit,
  L., Fortin, D., Cunanne, S., Descoteaux, M.:
\newblock Recognition of white matter bundles using local and global
  streamline-based registration and clustering.
\newblock NeuroImage \textbf{170} (2018)  283--295

\end{thebibliography}

\end{document}